\documentclass{JHEP3}

\usepackage{graphicx}
\usepackage{verbatim}
\usepackage{epsf}
\usepackage{latexsym}
\usepackage{amsmath,amsfonts,amssymb,amsthm}

\def\bseq{\begin{subequation}}  
\def\eseq{\end{subequation}}
\def\bsea{\begin{subeqnarray}}  
\def\esea{\end{subeqnarray}}

\newcommand{\bbox}{\lower.2ex\hbox{$\Box$}}

\newcommand{\beq}{\begin{equation}}
\newcommand{\eeq}{\end{equation}}
\newcommand{\bea}{\begin{eqnarray}}
\newcommand{\eea}{\end{eqnarray}}
\newcommand{\ena}{\end{eqnarray}}

\newcommand {\non}{\nonumber}

\newcommand{\be}{\begin{equation}}
\newcommand{\ee}{\end{equation}}

\title{\begin{center} 
Two Loop $\boldsymbol{R}$-Symmetry Breaking 
\end{center}}
\author{Antonio Amariti$^{1,a}$ and Alberto Mariotti$^{2,b}$ 
\\ ~
\\
$^1$Dipartimento di Fisica, Universit\`a di Milano Bicocca\\
and \\
INFN, Sezione di Milano-Bicocca,\\ 
piazza della Scienza 3, I-20126 Milano, Italy\\
\\
$^2$
Theoretische Natuurkunde, Vrije Universiteit Brussel \\
and\\
The International Solvay Institutes\\ 
Pleinlaan 2, B-1050 Brussels, Belgium\\
 ~~\\
  $^a$\email{antonio.amariti@mib.infn.it} ~~~
$^c$\email{alberto.mariotti@vub.ac.be}
}
\abstract{
We analyze two loop quantum corrections for pseudomoduli 
in O'Raifeartaigh like models.
We argue that $R$-symmetry can be spontaneously broken 
at two loop in non supersymmetric vacua.
We provide a basic example 
with this property.
We discuss on phenomenological applications.
}

\begin{document}

\section{Introduction}

In the last years a lot of effort has been devoted to the study of
supersymmetry breaking in metastable vacua.  The starting point has
been to understand \cite{ISS} that metastable and dynamical
supersymmetry breaking is a generic phenomenon in $\mathcal{N}=1$
gauge theories \cite{ISS,Main_examples,antofigo}. Subsequently these
classes of models have been studied in phenomenological settings,
where they have been considered as hidden sectors in gauge
mediation scenarios \cite{GAUGEMED,oogurifigo}.

In this framework $R$-symmetry breaking (\cite{shih}-\cite{RSYMBR})
plays an important role.  In this note we propose a mechanism for
spontaneous $R$-symmetry breaking in supersymmetry breaking vacua
through two loop effects.

In the next section we review some aspects
about $R$-symmetry
breaking in model building.  In section 2 we survey a
class of models and the corresponding two loop effective potential.  In
section 3 we present the basic model of metastable supersymmetry
breaking with two loop $R$-symmetry breaking.
In section 4 we conclude and comment.  In the Appendix \ref{Details2}
we give the details of the two loop computation and in the Appendix
\ref{appquiv} we embed our model in a quiver gauge theory.

\subsection{$R$-symmetry breaking}

Soft supersymmetry breaking \cite{GiraGrisa} in the 
MSSM is induced by a  mediation mechanism. 
This is necessary since
the supertrace theorem \cite{FGP}
\be 
\text{STr} \mathcal{M}^2=\sum_{J}(-1)^{2J}(2J+1)M_J^2 
\ee
constraints at tree level some masses of the superpartners to
be lower than the masses of the ordinary  particles.

In the case of gauge mediation, supersymmetry 
is broken in a secluded sector,
and then transmitted to the visible sector (the MSSM).
The breaking in the hidden sector can be spontaneous or
dynamical, even in metastable vacua.
%
There is a messenger sector 
coupled with the secluded sector and charged
under the MSSM gauge symmetries.
Loops of gauge fields and messengers
transmit the supersymmetry breaking to the MSSM.
In this way
the gauginos and the 
scalar superpartners of the ordinary fermions get
soft masses. However,
gaugino mass arises from loop with the messengers only
if $R$-symmetry is 
broken.
This is true if gauginos have Majorana mass.
If gauginos have Dirac mass,
it could be generated also in an
$R$-symmetry preserving model \cite{Dirac}.
We focus here on models with gaugino
Majorana mass.

The candidate model for the hidden sector should break both
supersymmetry and $R$-symmetry.
This is not easy because of a result of \cite{NelsonSeiberg}.
Models with explicit $R$-symmetry breaking and supersymmetry breaking
has been studied in \cite{antofigo,oogurifigo,Noietal}.
On the other hand, the possibility of spontaneous $R$-symmetry
breaking is quite natural.
A relevant issue here is the fate of the Goldstone boson of the
$R$-breaking.  
To evade
astrophysical constraints, this $R$-axion has to be massive, 
see \cite{BRP}.  
The axion can get mass by an
explicit breaking of $R$-symmetry, in a sector that does not couple at
tree level with the supersymmetry breaking sector.  This can be
realized through higher dimensional operators (suppressed by $1/M_p$)
or by the cancellation of the cosmological constant from supergravity.

Spontaneous $R$-symmetry breaking 
at tree level or at quantum level
has
been studied in O'Raifeartaigh models.  
In \cite{shih} it was shown that
$R$-symmetry can be broken at one loop if some of the fields of the
model have charge different from $0$ and $2$.  In \cite{sun} a similar
result was derived for tree level spontaneous $R$-symmetry
breaking.
In a recent paper \cite{GKKS} two loop corrections 
were shown to destabilize
an $R$-charged field at the origin of the pseudomoduli
space.  Then the addition of a small tree level effect stabilize this
field in the large vev region, breaking $R$-symmetry.

Here we look for models with spontaneous $R$-symmetry breaking at two
loop.  This breaking occurs when an $R$-charged field gets a vev only
from the two loop effective potential.
We show that
different couplings in the superpotential
lead to different signs for
the two loop mass.
The strategy is to combine these
contributions to give non zero vev
to the $R$-charged
field.

\section{One loop flat directions}

In this section we present the class of models we consider
through the paper. They are theories of pure chiral fields
with canonical Kahler potential and with
a renormalizable superpotential.
We study two loop corrections in models
with spontaneous breaking of supersymmetry. The
most natural way consists in coupling an
O'Raifeartaigh sector to another bunch of fields
through trilinear couplings. This implies that the
one loop corrections lift the potential for
the O'Raifeartaigh field but do not lift 
pseudomoduli space of the other sector.
The superpotentials we consider are
\be 
\label{class}
W = h
\left(
- \frac{f}{2} X + X \phi_1^2 + \mu \phi_1 \phi_2 + \phi_I \rho \xi_J +
    Z \xi_4^2 + \mu \xi_4 \xi_5
\right)
\ee
where $I=1,2$ and $J=4,5$.

The supersymmetry breaking vacuum is at the origin of the moduli
space. The fields $\phi_2$ and $\xi_4$ and $\xi_5$ have positive
squared mass $h^2 \mu^2$.  The field $\phi_1$ splits its mass in $ h^2
\mu^2 \pm h^2 f$, for its real and imaginary component.  The other
fields are pseudomoduli.  The pseudomodulus $X$ is stabilized at one
loop at the origin.  The pseudomodulus $\rho$ is also stabilized at one
loop at the origin, when $I=1$, i.e. when it is directly coupled in
the trilinear term with the field $\phi_1$ which has a mass splitting.
For the case with $I=2$ we add a mass term for the field $\rho$, to
avoid tachyons
\be
\Delta W_{I=2}= m \rho^2
\ee
with $m \ll \mu$.

The pseudomodulus $Z$ is not lifted at one loop and a two loop analysis is
required.  In the Appendix \ref{Details2} we give the details of the
calculation.  We summarize in Table 1 the results for the two loop
mass of the field $Z$, at order $o(m)$ for the cases $I=2$.
\begin{table} 
\begin{center}
\begin{tabular}{c|c|c|c}
I&J&$m_Z^2$&\\
\hline
1&4&$h^6 \mu^2(\log{4}-1-\frac{\pi^2}{6})$&$<0$\\
\hline
1&5&$h^6 \mu^2(\log{4}-1)$&$>0$\\
\hline
2&4& $h^6 \mu^2(\log{2}-1+\frac{\pi^2}{12})$&$>0$\\
\hline
2&5& $h^6 \mu^2(\log{2}-1)$&$<0$\\
\end{tabular} 
\end{center}
\caption{Two loop squared mass for $Z$. }
\label{albeCULO}
\end{table}
The model $(I,J)=(1,4)$ gives the same result than \cite{GKK}.
In fact it is the same model of pure chiral fields.
Then, the explicit calculation shows that the model
with $(I,J)=(2,5)$ has a runaway behaviour, while
the two loop potential for the $Z$ field
in the models $(I,J)=(1,5)$ and $(I,J)=(2,4)$ has a stable
minimum at the origin and the potential increases 
in the large field region.

The model $(I,J)=(1,4)$ has the bad behaviour discussed in the
surveying of \cite{Ken}.  Methods of \cite{Ken} 
can be generalized for
the other three models as well.  
For the
case with $I=4$ the beta function of the mass term $\mu \phi_1 \phi_2$
has to be taken into account.  
Moreover, in the
cases with $J=5$, the field $\xi_5$ does not decouple at large $Z$, and
the term $\phi_I \rho \xi_J$ affects the effective potential for long
RG time (i.e. large $Z$).  
In all the cases this analysis gives the same
qualitative behaviour than our explicit computation\footnote{We 
are grateful to Ken Intriligator for explaining us how to
  analyze these cases with the techniques of \cite{Ken}.}.
 
One can also notice that under the exchange $\xi_4 \leftrightarrow \xi_5$
the quadratic mass for the $Z$ field changes sign. 
This change corresponds to an
opposite $R$-symmetry charge for the field $Z$. Connecting the
behaviour of the two loop potential for $Z$ with its $R$-charge is an
interesting question that we leave for future investigation.

\subsubsection*{Breaking $R$-symmetry at two  loop}
In Table \ref{albeCULO} we observe that masses of different signs are
related to different trilinear couplings between the $\phi_i$ and the
$\xi_i$ sector.  Combining these contributions we can generate a two
loop potential for the field $Z$ which stabilizes it, but not at the
origin.  In the following we study a model with several
trilinear couplings.  The field $Z$ acquires a non trivial vev $\langle
Z \rangle \neq 0$ in the true quantum minimum at two loop.  The model
has a tree level $R$-symmetry, and the field $Z$ has a non trivial
$R$-charge, then the $R$-symmetry is spontaneously broken by the two loop
corrections.

\section{The basic example}\label{globsymm}

In this section we present the model that breaks supersymmetry and
perturbatively $R$-symmetry at two loop.  The superpotential is
\be \label{Themodel}
W = -h \frac{f}{2}  X+ h X \phi_1^2 + h \mu \phi_1 \phi_2 + 
    h \alpha \phi_1 \phi_7 \phi_6 + h \beta \phi_1 \phi_8 \phi_5
    + h \gamma \phi_2 \phi_7 \phi_5 + h\ Z  \phi_4^2 + h \mu \phi_4 \phi_5
    + h \mu  \phi_{6}^2
\ee
where $h$ is a marginal coupling, and $\alpha$, $\beta$ and $\gamma$
are numerical constants. All the couplings can be made real with a
phase shift of the fields.

We give in Table \ref{U1RZ2}
the $R$ charges and the $Z_2$ discrete symmetries.
\begin{table}
\begin{center}
\begin{tabular}{c|c|c|c|c|c|c|c|c|c}
         & $X$ & $\phi_1$ & $\phi_2$ & $\phi_4$ & $\phi_5$ 
       & $\phi_6$ & $\phi_7$ 
&$\phi_8$ & Z \\
\hline
$U(1)_R$ & 2 & 0 & 2 & 3 & -1 & 1  & 1  & 3  & -4 \\
\hline
$Z_2$ & 0 & $\pi$ & $\pi$ & 0 & 0& 0 & $\pi$ & $\pi$ & 0\\
\hline
$Z_2$ & 0 & $\pi$ & $\pi$ & $\pi$ & $\pi$ & $\pi$ & 0 & 0 & 0\\
\hline
$Z_2$ & 0 & 0 & 0 & $\pi$ & $\pi$& $\pi$ & $\pi$ & $\pi$ & 0\\
\end{tabular}
\end{center}
\caption{$U(1)_R$ and $Z_2$ charges of the fields.}
\label{U1RZ2}
\end{table}
These global symmetries and renormalizability constraints the theory to
the form (\ref{Themodel}), except for three terms $Z \phi_8^2$,
$\mu \phi_7^2$, $X\phi_5 \phi_6$.  In the limit $f \to 0$ the
theory admits a $U(1)$ global symmetry which forbids the terms $Z
\phi_8^2$, $\mu \phi_7^2$.  The term $X \phi_5 \phi_6$ has to be
tuned to zero.  It cannot be forbidden even introducing global
symmetries involving the couplings, to be thought
as spurion fields.  
A possible solution to this tuning is discussed
in the conclusion.

There is a supersymmetry breaking
vacuum at the origin of the moduli space.  Around this vacuum the
fields $\phi_2$, $\phi_4$, $\phi_5$ and $\phi_6$ have positive squared
mass $h^2 \mu^2$.  The field $\phi_1$ splits its mass in $ h^2 \mu^2
\pm h^2 f$, 
which are both
positive for $y=|f/\mu^2|<1$. The other fields are pseudomoduli, and
their squared mass spectrum has to be analyzed by looking at the loop
expansion of the scalar potential.

\subsection{One loop corrections}
The one loop corrections lift the $X$ and $\phi_8$ directions and set
$\langle X \rangle =0$ and $\langle \phi_8 \rangle =0$, with positive    
squared masses
\bea
&&m_X^2=\frac{16 h^4}{f} \left(
-2 \mu^2 f - 8 \mu^2 f \log[ h \mu] - (\mu^2-f)^2 \log [h^2 (\mu^2-f)]+ 
(\mu^2+f)^2 \log [h^2 (\mu^2+f)]
\right) \non
 \\
&&m_{\phi_8}^2= \frac{|\beta|^2}{4} m_X^2
\eea

\begin{figure}
\begin{center}
\includegraphics[width=10cm]{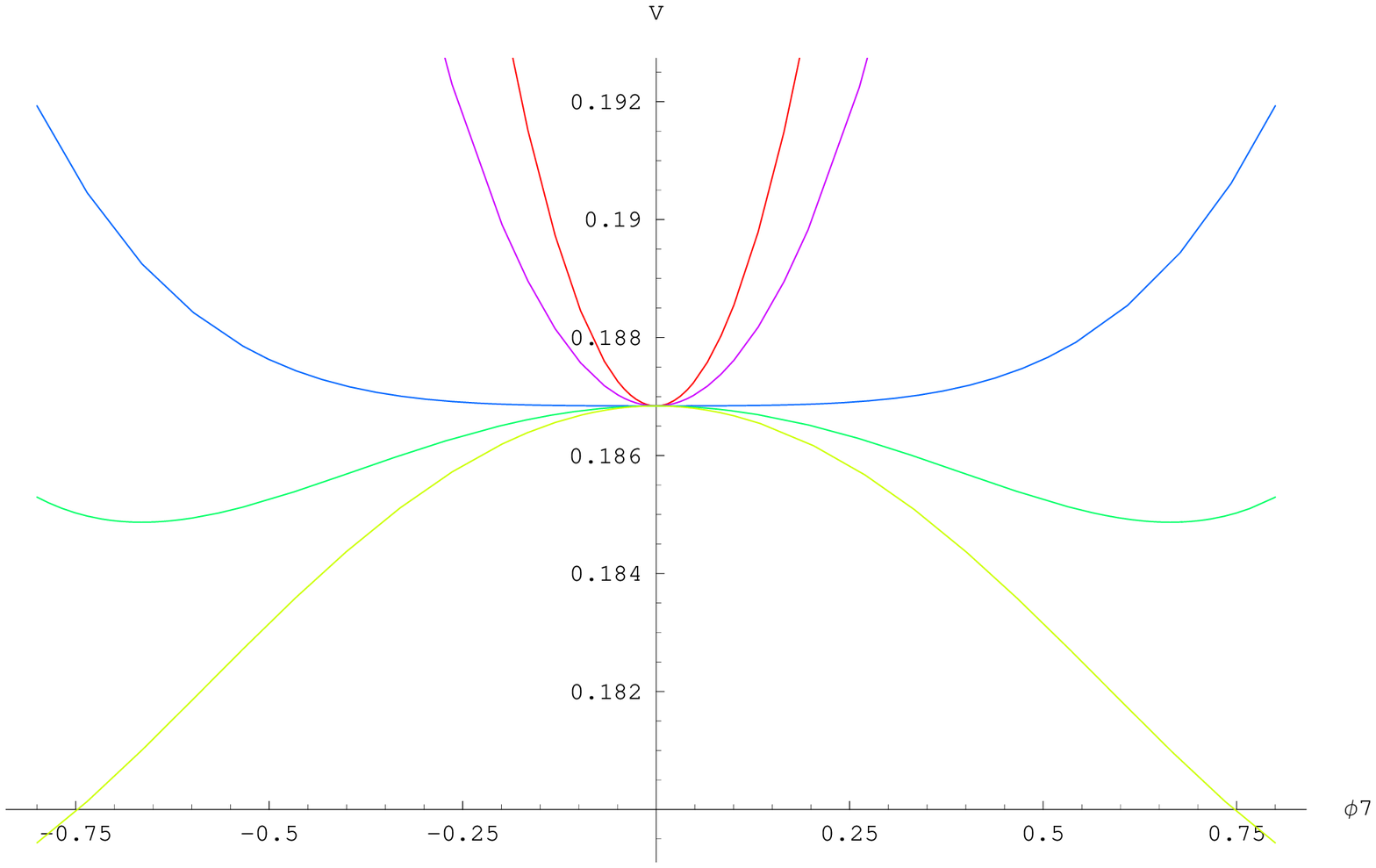}
\caption{Scalar potential for $\phi_7$ at the origin.  The ratio
  $\alpha/\gamma$ is respectively $2$, $1$, $0.9$, $0.8$ and $0.7$
  from the red to the yellow curve.}\label{FIGURA1}
\end{center}
\end{figure}

The fields $\phi_7$ is also stabilized at the origin but this
direction can develop a runaway behaviour to be analyzed.  First note
that this pseudomoduli space is stable for
\be
\label{rangepip}
|\langle \phi_7 \rangle |< \frac{\mu}{\gamma}\sqrt{\frac{1-y}{y}}
\ee
Figure \ref{FIGURA1} then shows for which values of the ratio
$\alpha/\gamma$ the one loop mass of $\phi_7$ is positive, after
fixing $f/\mu^2 = 0.5$ (all the other choices with $y<1$ are
possible).  We choose the ratio $\alpha/\gamma$ to stabilize the field
$\phi_7$ at the origin.

For $\phi_7$ larger than
(\ref{rangepip}) the theory has a runaway behaviour, parametrized by
$\phi_7 \rightarrow \infty$,
\bea 
&&
\phi_1 \sim \sqrt{\frac{f}{2}}
\ \ \ \ \
\phi_{2} \sim \sqrt{\frac{f}{2}}\frac{\alpha^2}{2 \gamma \mu^2}\phi_7^2
\ \ \ \ \ 
\phi_4 \sim 0
\ \ \ \ \ 
\phi_ 5 \sim \sqrt{\frac{f}{2}}\frac{\mu}{\gamma} \frac{1}{\phi_7}
\ \ \ \ \ 
\phi_6 \sim \sqrt{\frac{f}{2}} \frac{\alpha }{2 \mu}\phi_7
\nonumber \\
&&
\phi_8 \sim \frac{\alpha^2}{2 \beta \mu^2}\phi_7^3
\ \ \ \ \
X \sim \frac{\alpha^2}{2 \gamma \mu} \phi_7^2
\ \ \ \ \
Z \sim \sqrt{\frac{f}{2}}\frac{ \mu^2}{\gamma} \frac{1}{\phi_4 \phi_7}
\eea

\subsection{Two loop corrections}
\begin{figure}
\begin{center}
\begin{tabular}{c}
\includegraphics[width=6cm]{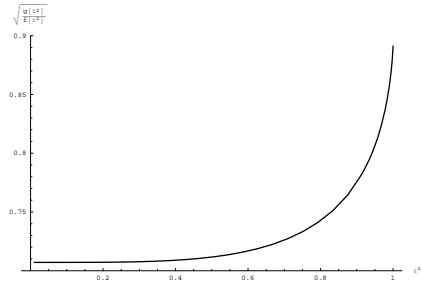}
\end{tabular}
\end{center}
\caption{Function $\sqrt{\frac{g(\tau^2)}{ f(\tau^2)}}$. 
The region of interest is the one below the curve.}
 \label{betagamma}
\end{figure}

\begin{figure}
\begin{center}
\includegraphics[width=15cm]{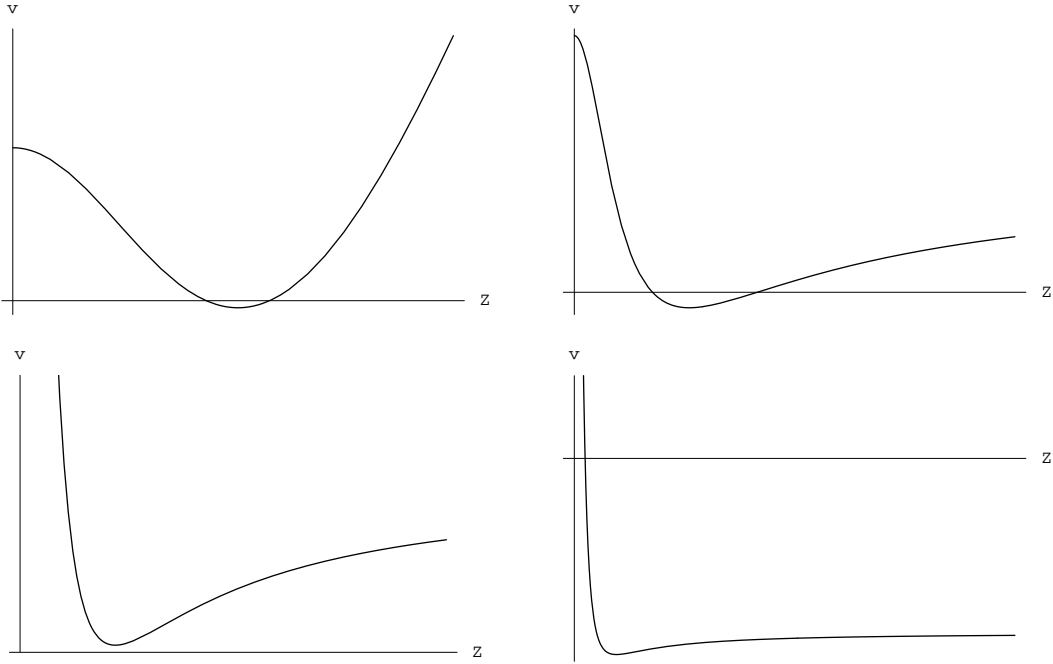}
\caption{Scalar potential for $Z$, plotted for different values of the
  ratio $\beta/\gamma$, respectively $0.7$, $0.65$, $0.6$ and $0.55$,
from left to right.
  The ration $f/\mu^2$ has been chosen to be $0.5$}\label{PotY}
\end{center}
\end{figure}

The potential for the field $Z$ 
is not lifted
at one loop, and a two loop
analysis is necessary. Considering $Z$ as a background field, the
masses of $\phi_4$ and $\phi_5$ mix.  We diagonalize the
fermionic mass matrix for these two fields.  The rotation is
\bea \label{mixing}
&&
\phi_4 = -s\theta \rho_4 + c\theta \rho_5
\nonumber \\
&&
\phi_5 = ~~c\theta \rho_4 + s\theta \rho_5
\eea
where 
\be
s\theta^2 = \frac{h^2\mu^2 - \lambda^2_{-}}{\lambda^2_{+}-\lambda^2_{-}}
\ee
and 
\be
\lambda^2_{\mp}=\frac{h^2}{2}\left(Z^2 + 2 \mu^2\mp Z\sqrt{Z^2+4\mu^2} \right)
\ee
The contributions 
to the two
loop effective potential for $Z$
are computed with the same strategy of \cite{GKK},
which is reviewed in Appendix \ref{Details2}.
The three contributions are given in Figure \ref{feyfig} and are
called $V_{SS}$, $V_{SSS}$ and $V_{FFS}$. We found
\bea
V_{SS}&=& h^2 \beta^2 c\theta^2
\left(
f_{SS}(h^2 (\mu^2-f), \lambda^2_{-})
+
f_{SS}(h^2 (\mu^2+f), \lambda^2_{-}) 
-
2 
f_{SS}(h^2 \mu^2, \lambda^2_{-})
\right)
\nonumber \\&&
+
\left(
c\theta^2,
\lambda_{-}
\leftrightarrow
s\theta^2,
\lambda_{+}
\right)
\nonumber \\
\nonumber \\
V_{SSS}&=&
h^4 \mu^2
( s\theta^2 \beta^2 + c\theta^2 \gamma^2)
(    f_{SSS}(0,h^2(\mu^2-f), \lambda^2_{-})+
     f_{SSS}(0,h^2(\mu^2+f), \lambda^2_{-})
\nonumber \\
&&
 -2 f_{FSS}(0,h^2\mu^2,   \lambda^2_{-})
)
+
\left(
s\theta^2,
c\theta^2,
\lambda_{-}
\leftrightarrow
c\theta^2,
s\theta^2,
\lambda_{+}
\right)
\nonumber \\
\nonumber \\
V_{FFS}&=&h^2 (\beta^2 c\theta^2)
(
f_{FSS}(0, \lambda^2_{-},h^2(\mu^2-f))+
f_{FSS}(0, \lambda^2_{-},h^2(\mu^2+f)) 
\nonumber \\
&&-2 f_{FSS}(0, \lambda^2_{-},h^2\mu^2)
)
+
\left( c\theta^2,\lambda_{-}
\leftrightarrow
s\theta^2,\lambda_{+}\right) 
\eea
Expanding the two loop effective potential for 
small $Z$, the mass term at the origin is
\be
\label{massaZ}
 m_Z^2 = h^6 \mu^2 (\beta^2 f(\tau^2) -
\gamma^2 g(\tau^2)) 
\ee 
where 
\bea 
\tau^2&&=f/\mu^2 \\
f(x) &&= -2 - \frac{(1-x)^2}{x}
\log(1-x) +\frac{(1+x)^2}{x} \log(1+x)
\nonumber\\
g(x) &&= 2 + \frac{(1-x)}{x} \log(1-x)- \frac{(1+x)}{x} \log(1+x) 
\eea
and the functions $f(x)$ and $g(x)$ are positive for $x<1$.
There is a regime of the parameters for which this mass term is
negative. 
This happens in the region
where 
$$\frac{\beta}{\gamma} < \sqrt{\frac{g(\tau^2)}{ f(\tau^2)}}$$
as in Figure \ref{betagamma}. 
In such a regime of the parameter we look for a minimum of the two loop scalar
potential.
We indeed observe in Figure \ref{PotY}, by plotting the scalar
potential, that there is a choice of the ratio $\beta/\gamma$ where
the scalar potential has a minimum at $\langle Z \rangle \neq 0$.

We can then conclude that the model (\ref{Themodel})
spontaneously breaks $R$-symmetry 
at two loop in a 
non supersymmetric (metastable) vacuum.
All the tree level flat
directions of the scalar potential
are lifted by quantum effects.
The vacuum is metastable because the field
$\phi_7$, which acquires positive squared mass 
around the origin through one loop corrections,
develops a runaway in the large vev region.
The effective potential for $\phi_7$ has to be
analyzed to estimate the lifetime of the
vacuum.

\section{Conclusions}

In this note we found a supersymmetry breaking model in which
$R$-symmetry is spontaneously broken at two loop in the scalar
potential. It is a model of pure chiral fields without any gauge
symmetry.  There is a tuning in the superpotential, since we did not
consider all the terms invariant under the global symmetries of the
theory. Adding the allowed term should spoil some of the infrared
properties, i.e. supersymmetry breaking.

The tuning problem can be solved by embedding the superpotential in a
quiver gauge theory, see Appendix \ref{appquiv}.  In this case the
pure chiral fields model has to be considered as the effective theory
around the non supersymmetric vacuum found at tree level in the gauge
theory, as in \cite{ISS}.  This embedding might also stabilize the
runaway behavior in the large field region, where strong dynamics
effects of the gauge groups add non perturbative terms to the
superpotential.


Moreover, this two loop analysis can be applied to many models with
metastable vacua. In most of them an approximate $R$-symmetry exists
at such vacua.  Two loop effects can offer a solution for this
problem.
Indeed, as in the model we studied, we can couple the theory to an
$R$-charged pseudomodulus that receives two loop corrections from the
supersymmetry breaking sector.  This field can acquire a quantum
scalar potential that breaks spontaneously $R$-symmetry.

Another possibility is to build a model with a ``tension'' between the one
loop and the two loop contributions for some pseudomoduli. This
competition could shift the minimum from the origin, breaking
$R$-symmetry.  In \cite{GKK} the one and two loop corrections in the
ISS model with a mass hierarchy among the fundamental fields have been
studied.  However in that case one can check that the quantum
corrections lead to a runaway, without any local minimum, and then
restore supersymmetry.  It would be interesting to find models where
the combination of one loop and two loop quantum corrections lead to
metastable vacua.
\section*{Acknowledgments}
We first thank Luciano Girardello for early collaboration on this
project, for many interesting discussions 
and for comments on the draft.
We would also like to thank Zohar Komargodski for useful comments, and
especially Ken Intriligator for helpful explanations.
A.~A.~ is supported in part by INFN
and
MIUR under contract 2007-5ATT78-002.
A. ~M. ~ is
supported in part 
by the Belgian Federal Science Policy Office 
through the Interuniversity 
Attraction Pole IAP VI/11, by the European
Commission FP6 RTN programme MRTN-CT-2004-005104 and by 
FWO-Vlaanderen through project G.0428.06.


\appendix

\section{Two loop effective potential}\label{Details2}

The calculation of the two loop effective potential for a pseudomodulus 
is involved, since a lot of graphs can contribute.
Here we used the trick of \cite{GKK}, which makes the calculation
simpler.  One has to switch off the supersymmetry breaking scale, $f$,
and compute the supersymmetric masses for all the fields. The
pseudomoduli are massless also in this supersymmetric version of
the model, but in this case they cannot be lifted by quantum
corrections.  The two loop potential for these fields can be calculated
by subtracting the supersymmetric part to the non supersymmetric one.
In formulas, calling $V^{(2)}(Z)$ the two loop potential, it is given
by
\be
V^{(2)}(Z) = V_{nonSUSY}^{(2)}(Z)-V_{SUSY}^{(2)}(Z) 
\ee
This formula means that the effective potential for $Z$ is 
due to the diagrams that depends both on the fields whose masses split in
the non supersymmetric case (to respect to the supersymmetric one)
and on the fields whose masses depend on Z.

There are only few diagrams of this form
and they are computed using the formulas of \cite{Jones,Martin}.
\begin{figure}
\begin{center}
 \includegraphics[width=10cm]{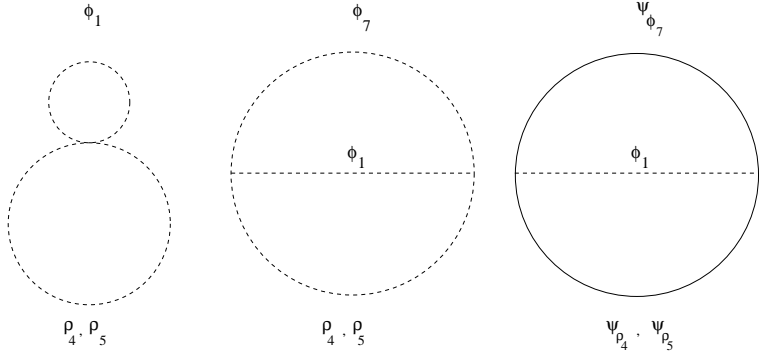}
\caption{Relevant Feynman graphs to the two loop potential}
\label{feyfig}
\end{center}
\end{figure}
In this paper, the model (\ref{Themodel}) gives rise to three 
different diagrams, $V_{SS}$, $V_{SSS}$ and $V_{FFS}$, and
they are given in Figure \ref{feyfig}.

\subsection{Details on the calculation}
Here we explain the details on the computation
of the mass for the pseudomodulus  $Z$ around the origin.
The potential is made of three different pieces
\be
V^{(2)}(Z) = V_{SS}+V_{SSS}+V_{FFS}
\ee 
They come from three different Feynman graphs, and they have
been explicitly derived in \cite{Jones,Martin}.
They are
\bea
&&V_{SS}(x,y)   = J(x)J(y)  \nonumber \\
&&V_{SSS}(x,y,x)= -I(x,y,z)  \nonumber \\
&&V_{FFS}(x,y,z)= J(x)J(y) - J(x)J(z) -J(y)J(z)+(x+y-z)I(x,y,z)
\eea
where
\be
J(x) = x \left(\log\frac{x}{Q^2}-1\right)
\ee
In our calculation one argument of the function $I(x,y,z)$ is
always zero. We give the expression for this simplified case
\bea
I(0,x,y) = &&(x-y) \left(
\text{Li}_2(y/x) - \log(x/y) \log\frac{x-y}{Q^2}
+
\frac{1}{2}\log^2\frac{x}{Q^2} -\frac{\pi^2}{6}
\right)
\nonumber \\
-&&\frac{5}{2} (x+y) + 2 x \log\frac{x}{Q^2} +2 y \log\frac{y}{Q^2}
-x \log\frac{x}{Q^2}\log\frac{y}{Q^2}
\eea
Using these formulas we found in (\ref{massaZ}) that the mass term for $Z$ is
$m_Z^2 = m_{Z_{\beta}}^2 +  m_{Z_{\gamma}}^2$, with $\tau^2 = \frac{f}{\mu^2}$, 
\bea \label{massbeta}
m_{Z_{\beta}}^2&& 
= \frac{h^6 \beta^2 \mu^2}{\tau^2}\left(
\phantom{\frac{\frac{a}{b}}{\frac{c}{c}}} \! \!\! \!\! \!
-2 \tau^2- (1-\tau^2)^2
\log(1-\tau^2) +(1+\tau^2)^2 \log(1+\tau^2)
\right.\nonumber \\
+&& \left.  \frac{1}{2}\log^2{(1+\tau^2)}+Li_{2}(-\tau^2)+
Li_{2}\left(\frac{\tau^2}{1+\tau^2} \right)\right)
\eea
and
\be
m_{Z_{\gamma}}^2 
= -\frac{h^6 \gamma^2 \mu^2}{\tau^2}
(2 \tau^2 +(1-\tau^2)\log(1-\tau^2)
-(1+\tau^2)\log(1+\tau^2)))
\ee
The last line in (\ref{massbeta}) vanishes for $\tau^2<1$ 
because of an identity of dilogarithms.

\section{Embedding in quiver gauge theories} \label{appquiv}
\begin{figure}
\begin{center}
\includegraphics[width=5cm]{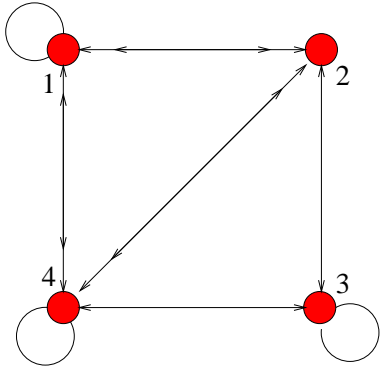}
\caption{Quiver for the superpotential (B.1). } 
\label{quiv}
\end{center}
\end{figure}
A possible embedding in quiver gauge theory is a four $U(1)$ nodes
theory (note that also non-abelian groups are admitted) with superpotential
\bea
W &&= q_{12}^{(1)} q_{21}^{(1)} X_{11} -f X_{11}
+ \mu \left(q_{12}^{(1)} q_{21}^{(2)}
+q_{12}^{(2)} q_{21}^{(1)} \right)+q_{21}^{(1)}q_{14}^{(7)} q_{42}^{(6)}
\nonumber \\
+&&q_{12}^{(1)}q_{24}^{(6)} q_{41}^{(7)}+q_{21}^{(2)}q_{14}^{(7)}q_{42}^{(5)}
+q_{12}^{(2)}q_{24}^{(5)}q_{41}^{(7)}+q_{21}^{(1)}q_{14}^{(8)}q_{42}^{(5)}
\nonumber \\
\label{supquiver}
+&&q_{12}^{(1)}q_{24}^{(5)}q_{41}^{(8)}+q_{23}^{(4)}Z_{33} q_{32}^{(4)}
+ q_{43}q_{32}^{(4)}q_{24}^{(5)} 
+ q_{42}^{(5)}q_{23}^{(4)} q_{34}   
+ q_{34} Y_{44} q_{43} - \mu^2 Y_{44}
\eea
The quiver is shown in Figure \ref{quiv}.
The upper scripts map the fields in (\ref{supquiver}) with
the corresponding fields in (\ref{Themodel}). 
The fields $q_{34}$ and $q_{43}$ get a vev $\mu$
from the equation of motion of the field $Y_{44}$.
This gives a mass term for the fields  $q_{32}^{(4)}q_{24}^{(5)}$
and $q_{42}^{(5)}q_{23}^{(4)}$, as in (\ref{Themodel}).

In this model the requirement of gauge invariance
forbids the dangerous term $X \phi_5 \phi_6$
that we discussed in section \ref{globsymm}.

\end{document}